\documentclass[12pt]{iopart}

\usepackage{iopams}
\usepackage{amsthm}
\usepackage{hyperref}

\newcommand{\Ai}{\mathrm{Ai}}
\renewcommand{\P}{\mathrm{Prob}}

\newtheorem{lemma}{Lemma}

\newcommand{\affiliation}[1]{\address{#1}}
\newcommand{\keywords}[1]{\vspace{10mm}\noindent\textbf{Keywords:} #1}
\newcommand{\tfrac}[2]{\mbox{\small$\frac{#1}{#2}$}}
\newcommand{\openone}{\mbox{{\small 1}$\!\!$1}}
\newcommand{\text}[1]{\mathrm{#1}}

\begin{document}

\title{The One-dimensional KPZ Equation and the Airy Process}
\author{Sylvain Prolhac\footnote{prolhac@ma.tum.de}{} and Herbert Spohn\footnote{spohn@ma.tum.de}{}}
\affiliation{Zentrum Mathematik and Physik Department,\\
Technische Universit\"at M\"unchen,\\
D-85747 Garching, Germany}
\date{\today}

\begin{abstract} Our previous work on the one-dimensional KPZ equation with sharp wedge initial data is extended to the case of the joint height statistics at $n$ spatial points for some common fixed time. Assuming a particular factorization, we compute an $n$-point generating function and write it in terms of a Fredholm determinant. For long times the generating function converges to a limit, which is established to be equivalent to the standard expression of the $n$-point distribution of the Airy process.
\end{abstract}

%02.30.Ik Integrable systems
%02.50.Cw Probability theory
%02.50.Ey Stochastic processes
%02.50.Ga Markov processes
%05.20.-y Classical statistical mechanics
%05.30.Jp Boson systems
%05.40.-a Fluctuation phenomena, random processes, noise, and Brownian motion
%05.70.Np Interface and surface thermodynamics

\maketitle

%%%%%%%%%%%%%%%%%%%%%%%%%%%%%%%%%%%%%
%%%%%%%%%%%%%%%%     %%%%%%%%%%%%%%%%
%%%%%%%%%%%%%%%%  1  %%%%%%%%%%%%%%%%
%%%%%%%%%%%%%%%%     %%%%%%%%%%%%%%%%
%%%%%%%%%%%%%%%%%%%%%%%%%%%%%%%%%%%%%

\section{Introduction}
\label{Section introduction}
\setcounter{equation}{0}
The two items in the title are rooted in disjoint areas. The Airy process naturally belongs to random matrix theory, while the one-dimensional Kardar-Parisi-Zhang (KPZ) equation is the stochastic evolution for droplet growth in a thin film. The purpose of the introduction is to put forward the two players, starting with stochastic growth.

The KPZ equation \cite{KPZ86.1} describes the evolution of the height function $h(x,t)$ on the line and reads
\begin{equation}
\label{KPZ}
\partial_{t}h(x,t)=\tfrac{1}{2}\lambda(\partial_{x}h(x,t))^{2}+\nu\,\partial_{x}^{2}h(x,t)+\sqrt{D}\,\eta(x,t)\;.
\end{equation}
Here the first term is the nonlinear growth velocity, to the lowest relevant order in the slope, and the Laplacian represents surface tension. The randomness of local attachments at the interface is modeled by Gaussian white noise, $\eta$, with mean zero and covariance
\begin{equation}
\label{<etaeta>}
\langle\eta(x,t)\eta(x',t')\rangle=\delta(x-x')\delta(t-t')\;.
\end{equation}
$\sqrt{D}$ is the noise strength.

As an evolution equation, (\ref{KPZ}) has to be supplemented with initial conditions. We are interested in droplet growth which is triggered by the sharp wedge initial condition
\begin{equation}
\label{KPZ t=0}
h(x,0)=-|x|/\epsilon\;
\end{equation}
in the limit $\epsilon\to0$. 
The solution to the KPZ equation is then of the form \cite{ACQ10,SS10.1,SS10.2}
\begin{equation}
\label{height scaling KPZ}
\gamma^{-1}h(x,t)=-\frac{\rho t}{24}-\frac{(\mu x)^{2}}{2\rho t}+\log\mu+\xi(x,t)
\end{equation}
with parameters 
\begin{equation}
\label{gamma mu rho}
\gamma=\frac{2\nu}{\lambda} \;,\quad \mu=\frac{\lambda^{2}D}{(2\nu)^{3}} \quad\text{and}\quad \rho=\frac{\lambda^{4}D^{2}}{(2\nu)^{5}}\;.
\end{equation}
Note that the random amplitude $\xi(x,t)$ is stationary in $x$ for fixed $t$. The KPZ theory predicts $\xi(x,t)$ to be of order $t^{1/3}$ and to have correlations in $x$ over a range of order $t^{2/3}$.

To proceed to the second player, the Airy process is most naturally introduced through Dyson's Brownian motion. Let us consider the dynamics of an $N\times N$ complex hermitian matrix, $A(t)$, as governed by the Langevin equation
\begin{equation}
\label{Langevin W}
\frac{d}{dt}A(t)=-\frac{1}{N}A(t)+W(t)\;.
\end{equation}
Here $W(t)$ is unitarily invariant $N\times N$ matrix-valued Gaussian white noise, \textit{i.e.} $W(t)$ is an $N\times N$ matrix, $W(t)=W(t)^{*}$, such that in some fixed basis
\begin{eqnarray}
&&\langle W_{ij}(t)\rangle=0\nonumber\\
&&\langle W_{ij}(t)W_{i'j'}(t')\rangle=\delta_{ij'}\delta_{i'j}\delta(t-t')\;.
\end{eqnarray}
In particular, for any unitary matrix $U$ it holds
\begin{equation}
UW(t)U^{*}=W(t)
\end{equation}
in distribution. The Langevin equation (\ref{Langevin W}) has a stationary solution, which is given by the Gaussian unitary ensemble (GUE)
\begin{equation}
\label{measure GUE}
Z^{-1}\,\rme^{-\frac{1}{N}\tr A^{2}}\mathrm{d}A\;.
\end{equation}
If one picks (\ref{measure GUE}) as initial distribution, then $A(t)$ is a stationary matrix-valued stochastic process. Some given trajectory in matrix space induces the motion of the eigenvalues of $A(t)$. They can be ordered as
\begin{equation}
\lambda_{1}(t)<\ldots<\lambda_{N}(t)\;,
\end{equation}
since with probability one there is no crossing. We are interested in the largest eigenvalue $\lambda_{N}(t)$. It is roughly located at the band edge $2N$. Rescaling relative to the edge, one has
\begin{equation}
\label{Airy process}
\lim_{N\to\infty}N^{-1/3}(\lambda_{N}(N^{2/3}w)-2N)=\mathcal{A}(w)\;,
\end{equation}
which we regard as the definition of the Airy process $\mathcal{A}(w)$. Since $\lambda_{N}(t)$ is stationary, so is the limit $\mathcal{A}(w)$.

(\ref{Airy process}) is a somewhat indirect prescription. A more direct characterization comes from the fact that the set of trajectories $\{\lambda_{j}(t),t\in\mathbb{R},j=1,\ldots,N\}$ form a determinantal process. Thus the finite dimensional distributions of $\mathcal{A}(w)$ can be written as a Fredholm determinant. At fixed $w$ the relevant operator is the Airy operator $K$ with the integral kernel
\begin{equation}
\label{K}
K(x,y)=\int_{0}^{\infty}\rmd u\,\Ai(u+x)\Ai(u+y)\;,
\end{equation}
called Airy kernel, where $\Ai$ is the standard Airy function. The transfer operator along the $w$-axis is governed by the Airy Hamiltonian
\begin{equation}
\label{H}
H=-\frac{d^{2}}{du^{2}}+u\;.
\end{equation}
Note that $K$ is the projection onto all negative energy eigenstates of $H$, thus $[H,K]=0$ and $K^2 = K$. In particular one can write $K(x,y)=\langle\Ai_{x}|K|\Ai_{y}\rangle$, where $\Ai_{x}(u)=\Ai(u+x)$ is the eigenfunction of $H$ with eigenvalue $-x$. Now let us pick times $w_{1}<\ldots<w_{n}$ and real numbers $a_{1}$, \ldots, $a_{n}$. Then
\begin{equation}
\label{n point proba Airy process}
\P\big(\mathcal{A}(w_{j})\leq a_{j},j=1,\ldots,n\big)=\det(1-\mathsf{K})\;.
\end{equation}
$\mathsf{K}$ is a $n\times n$ matrix with operator entries, 
\begin{eqnarray}\label{matrix kernel}
\hspace{-40pt}\langle x|\mathsf{K}_{ij}|y\rangle = \left\{
\begin{array}{rl}
 \openone_{\{x\geq a_{i}\}} \langle x|\rme^{(w_{i}-w_{j})H}K|y\rangle \openone_{\{y\geq a_{j}\}}&\mathrm{for}\quad i\geq j\;,\\
 - \openone_{\{x\geq a_{i}\}} \langle x|\rme^{(w_{i}-w_{j})H}(1 - K)|y\rangle \openone_{\{y\geq a_{j}\}} & \mathrm{for} \quad i < j \;.
\end{array}\right.
\end{eqnarray}
 $\mathsf{K}_{ij}$ is trace class and the Fredholm determinant (\ref{n point proba Airy process}) is well defined. The marginal distribution at one single point equals $\det (1 - P_{a}KP_{a})$ with $P_{a}$ the projection operator onto the interval $[a,\infty)$, which one recognizes as the GUE Tracy-Widom distribution function.

One knows that the Airy process is almost surely continuous and looks locally like a Brownian motion. In particular for the covariance, $\langle(\mathcal{A}(w)-\mathcal{A}(0))^{2}\rangle=|w|$ for $|w|\to0$ and
\begin{equation}
\langle\mathcal{A}(0)\mathcal{A}(w)\rangle-\langle\mathcal{A}(0)\rangle\langle\mathcal{A}(w)\rangle\simeq\frac{2}{w^{2}}\;
\end{equation}
for large $|w|$. We refer to \cite{ST10} for more detailed information. For polynuclear growth \cite{PS02.2} and the single step model \cite{J03} it has been established that in the long time limit, on the scale $t^{1/3}$ for the height and $t^{2/3}$ in the transverse direction, the height statistics is well approximated by the Airy process. On this basis one expects the same property to hold for the KPZ equation. More specifically, for the fluctuation term in (\ref{height scaling KPZ}) it should hold
\begin{equation}
\label{scaling fluctuations KPZ}
\lim_{t\to\infty}t^{-1/3}\xi(2\mu t^{2/3}w,2t)=\mathcal{A}(w)\;,
\end{equation}
where the nonuniversal coefficient $\mu$, see (\ref{gamma mu rho}), results from the KPZ scaling theory \cite{K95}.

For a single point, $n=1$, the property (\ref{scaling fluctuations KPZ}) has recently been established. The results in \cite{ACQ10,SS10.1,SS10.2,SS10.3} rely on the approximation of the KPZ equation through the weakly asymmetric simple exclusion process. Alternatively one can use the replica method \cite{K87.1,CLDR10.1,D10.1,D10.2}. The $m$-th replica is of order $\exp[tm^{3}/3]$ and one is forced to make an analytic continuation in $m$ into the complex 
plane $\mathbb{C}$.
At this point one adopts the choice
\begin{equation}
\label{1.17}
\exp[tm^{3}/3] = \int\rmd u\,\Ai(u)\rme^{umt^{1/3}}\;,
\end{equation}
which is supported by the fact that thereby the one-point generating function agrees with the one in \cite{ACQ10,SS10.1}.

For multi-point distributions, currently only the replica method seems to be available. We regard it as compelling that the analytic continuation should be chosen as in (\ref{1.17}). For two points, $w_{1}$, $w_{2}$, such a program is carried out in \cite{PS10.1}. Under a decoupling approximation it is established that $t^{-1/3}\xi(2\mu t^{2/3}w_{j},2t)$, $j=1,2$, jointly converge to $\mathcal{A}(w_{1}), \mathcal{A}(w_{2})$ as $t\to\infty$. In fact, one first computes the generating function
\begin{equation}
\label{GF 2pt}
\langle\exp(-\rme^{-s_{1}+\xi(x_{1},t)}-\rme^{-s_{2}+\xi(x_{2},t)})\rangle
\end{equation}
in terms of a Fredholm determinant, which is sufficiently explicit to easily read off the long time limit.

The goal of this contribution is to extend the results from \cite{PS10.1} to an arbitrary but finite number of points $x_{1}$, \ldots, $x_{n}$. In particular we will study the generating function 
\begin{equation}\label{GF npt}
\Big\langle\exp\Big(-\sum_{j=1}^{n} \rme^{-s_{j}+\xi(x_{j},t)}\Big)\Big\rangle\;.
\end{equation}
In the properly scaled limit $t\to\infty$ one obtains a Fredholm determinant of an operator acting in $L^2(\mathbb{R})$, which looks very different from the standard definition given above. In \ref{Appendix matrix kernel} the equality of both expressions will be established.

In the next section we recall the replica method based on the link of the KPZ equation to the continuum directed polymer. We also state our main result. In Section \ref{Section replica summation} we carry out the replica summation, which is used in Section \ref{Section Fredholm G1} to compute the generating function (\ref{GF npt}).

%%%%%%%%%%%%%%%%%%%%%%%%%%%%%%%%%%%%%
%%%%%%%%%%%%%%%%     %%%%%%%%%%%%%%%%
%%%%%%%%%%%%%%%%  2  %%%%%%%%%%%%%%%%
%%%%%%%%%%%%%%%%     %%%%%%%%%%%%%%%%
%%%%%%%%%%%%%%%%%%%%%%%%%%%%%%%%%%%%%

\section{Replica approach and main result}
\label{Section replica}
\setcounter{equation}{0}
Through the Cole-Hopf transformation,
\begin{equation}
\label{Z[h]}
Z(x,t)=\rme^{\frac{\lambda}{2\nu}h(x,t)}\;,
\end{equation}
the KPZ equation becomes
\begin{equation}
\label{Z}
\partial_{t}Z(x,t)=\nu\,\partial_{x}^{2}Z(x,t)+(\lambda\sqrt{D}/2\nu)\,\eta(x,t)Z(x,t)
\end{equation}
and the sharp wedge initial condition translates to
\begin{equation}
\label{Z[h]0}
Z(x,0)= \delta(x)\;.
\end{equation}
The solution of (\ref{Z}), (\ref{Z[h]0}) can be written through the Feynman-Kac formula as
\begin{equation}
\label{Z(x,t)}
\fl\; Z_{\lambda,\nu,D}(x,t)=\int_{x(0)=0}^{x(t)=x}\mathcal{D}[x(\tau)]\,\exp\left({-\int_{0}^{t}\rmd\tau\left[\tfrac{1}{4\nu}(\partial_{\tau}x(\tau))^{2}-\tfrac{\lambda\sqrt{D}}{2\nu}\eta(x(\tau),\tau)\right]}\right)\;.
\end{equation}
The average is over all Brownian motion paths starting at 0 at time $t$ and ending at $x$ at time $t$. Thus $Z(x,t)$ can be viewed as the partition function of a continuum directed polymer model in $1+1$ dimensions immersed in a white noise random potential.
The precise definition of (\ref{Z(x,t)}) requires some discussion. We refer to \cite{SS10.4} for details. After energy renormalization
the average partition function is given by
\begin{equation}
\label{<Z>}
\langle Z_{\lambda,\nu,D}(x,t)\rangle=\frac{1}{\sqrt{4\pi\nu t}}\,\exp\left(-\tfrac{x^{2}}{4\nu t}\right)\;.
\end{equation}

From the scale invariance of Brownian motion and of white noise one concludes that 
\begin{equation}
Z_{\lambda,\nu,D}(x,t)=\mu\,Z_{1,1/2,1}\left(\mu\,x,\rho\,t\right)
\end{equation}
with the notations of (\ref{gamma mu rho}). Thus it suffices to study a particular set of parameters and, from now on, we set
\begin{equation}
\label{set parameters}
\lambda=1\;,\quad\nu=\tfrac{1}{2}\;,\quad D=1\quad\text{and}\quad Z_{1,1/2,1}(x,t)=Z(x,t)\;.
\end{equation}
More important in our context is the observation that the ratio $Z(x,t)/\langle Z(x,t)\rangle$ for fixed $t$ is stationary stochastic process in $x$. In particular the $n$-point function is invariant after shifting all arguments by the same fixed amount. On the level of the KPZ equation this means that under the parabolic shift, $h(x,t) + x^{2}/2t$, the statistical fluctuations of the interface become stationary in $x$.

In the replica approach the $n$-th moment of $Z$ is written as a matrix element of the propagator for the attractive $\delta$-Bose gas. We introduce the Lieb-Liniger Hamiltonian
\begin{equation}
\label{H Bose}
H_{n}=-\frac{1}{2}\sum_{i=1}^{n}(\partial_{x_{i}})^{2}-\frac{1}{2}\sum_{i\neq j=1}^{n}\delta(x_{i}-x_{j})\;.
\end{equation}
Then
\begin{equation}
\label{PDE <ZZZ>}
\left\langle Z(x_{1},t)\ldots Z(x_{n},t)\right\rangle= \langle 0,\ldots,0|\rme^{-H_n t}|x_{1},\ldots,x_{n}\rangle\;,
\end{equation}
where $|x_{1},\ldots,x_{n}\rangle$ is the quantum state with the particles at positions $x_{1},\ldots,x_{n}$. Since the initial condition is a symmetric function, the propagator is needed only in the symmetric subspace. On the level of the $\delta$-Bose gas, the energy renormalization corresponds precisely to the omission of the self-energy in (\ref{H Bose}). 

The quantity of interest is the generating function
\begin{equation}
\label{G[xi]}
G(\boldsymbol{s};\boldsymbol{x},t)=\Big\langle\exp\Big(-\sum_{j=1}^n \rme^{-s_{j}+\xi(x_{j},t)}\Big)\Big\rangle\;,
\end{equation}
setting $\boldsymbol{s}=(s_{1},\ldots,s_{n})$ and $\boldsymbol{x}=(x_{1},\ldots,x_{n})$ with $x_{1}<\ldots<x_{n}$. Note that according to (\ref{height scaling KPZ}) the term $-\tfrac{1}{2t}x^2 -\tfrac{1}{24}t$ has been subtracted from $h(x,t)$ already. To compute the generating function, one expands the exponential in (\ref{G[xi]}), writes the multinomials in $Z(t)$ as in (\ref{PDE <ZZZ>}), and expands the propagator in terms of the exact eigenstates of $H_n$ \cite{D10.2}. As in \cite{PS10.1}, for $n \geq 2$ this program can be carried out under a natural factorization assumption. To distinguish from the exact generating function, we add the superscript $^{\sharp}$ whenever the factorization is invoked. 

Our main result is a Fredholm determinant expression for $G^{\sharp}$. Setting 
$\alpha=(t/2)^{1/3}$, we obtain
\begin{equation}
\label{Gsharp[Q]}
G^{\sharp}(\boldsymbol{s};\boldsymbol{x},t)=\det(1- QK)\;,
\end{equation}
where $K$ is the Airy kernel (\ref{K}) and the integral kernel of $Q$ is defined by
\begin{eqnarray}
\label{Q}
&&\fl\quad \langle u_{1}|Q|u_{n+1}\rangle=\int_{-\infty}^{\infty}\rmd u_{2}\,\ldots\,\rmd u_{n}\,\langle u_{1}|\rme^{((x_{1}-x_{2})/{2\alpha^{2})}H}|u_{2}\rangle \ldots\langle u_{n}|\rme^{((x_{n}-x_{1})/2\alpha^{2})H}|u_{n+1}\rangle\nonumber\\
&&\fl\qquad\qquad\qquad\qquad\qquad\qquad\qquad\qquad\qquad \times\Phi(\alpha u_{1}-s_{1},\ldots,\alpha u_{n}-s_{n})\;,
\end{eqnarray}
with
\begin{equation}
\label{Phi}
\Phi(u_{1},\ldots,u_{n})=\frac{\rme^{u_{1}}+\ldots+\rme^{u_{n}}}{1+\rme^{u_{1}}+\ldots+\rme^{u_{n}}}\;.
\end{equation}

In the long time limit, the shifted height $\xi(x,t)$ scales as $ t^{1/3}$, and the joint distribution of the $\xi(x_{\ell},t)$'s is nontrivial if the distances are scaled as $t^{2/3}$. Let us thus substitute $s_{\ell}$ by $\alpha a_{\ell}$ and $x_{\ell}$ by $2\alpha^{2}y_{\ell}$. The right hand side of (\ref{G[xi]}) converges to the probability that $\xi(x_{j},t)$ is smaller than $\alpha a_{j}$ for all $j$, while $\Phi$ in (\ref{Q}) gives the characteristic function of the set $\{u_{j}\geq a_{j},j=1,\ldots,n\}$. Hence
\begin{equation}
\lim_{t\to\infty}\P\big(\xi(2\alpha^{2}y_{1},t)\leq\alpha a_{1},\dots,\xi(2\alpha^{2}y_{n},t)\leq\alpha a_{n}\big)=F_{n}(\boldsymbol{a},\boldsymbol{y})
\end{equation}
and
\begin{equation}
\label{2.26}
\fl\qquad\qquad F_{n}(\boldsymbol{a},\boldsymbol{y})=\det\big(1-K+\bar{P}_{a_{1}}\rme^{(y_{1}-y_{2})H}\bar{P}_{a_{2}}\rme^{(y_{2}-y_{3})H}\ldots\bar{P}_{a_{n}}\rme^{(y_{n}-y_{1})H}K\big)\;.
\end{equation}
Here $\bar{P}_{a}=1-P_{a}$ is the projector on $(-\infty,a]$, in other words $\langle u|\bar{P}_{a}|v\rangle=\openone_{\{u\leq a\}}\delta(u-v)$.
The agreement of (\ref{2.26}) with the standard definition (\ref{n point proba Airy process}) will be established in \ref{Appendix matrix kernel}.

%%%%%%%%%%%%%%%%%%%%%%%%%%%%%%%%%%%%%
%%%%%%%%%%%%%%%%     %%%%%%%%%%%%%%%%
%%%%%%%%%%%%%%%%  3  %%%%%%%%%%%%%%%%
%%%%%%%%%%%%%%%%     %%%%%%%%%%%%%%%%
%%%%%%%%%%%%%%%%%%%%%%%%%%%%%%%%%%%%%

\section{Replica Summation}
\label{Section replica summation}
\setcounter{equation}{0}
We start from the generating function $G$ of Eq. (\ref{G[xi]}), set
\begin{equation}
G_{1}(\boldsymbol{s};\boldsymbol{x},t)=\Big\langle\exp\big(-\sum_{\ell=1}^{n}\rme^{(t/24)-s_{\ell}}Z(x_{\ell},t)\Big)\Big\rangle =G(\boldsymbol{\tilde{s}};\boldsymbol{x},t)\;,
\end{equation}
where $\tilde{s}_{\ell}=s_{\ell}+x_{\ell}^{2}/2t$, and expand the exponential as
\begin{equation}
\label{G1[<ZZZ>]}
\fl\qquad G_{1}=1+\sum_{N=1}^{\infty}\frac{(-1)^{N}\rme^{tN/24}}{N!}\sum_{\ell_{1},\ldots,\ell_{N}=1}^{n}\rme^{-(s_{\ell_{1}}+\ldots+s_{\ell_{N}})}\left\langle Z(x_{\ell_{1}},t)\ldots Z(x_{\ell_{N}},t)\right\rangle\;.
\end{equation}
By (\ref{PDE <ZZZ>}) the $N$-point function of $Z$ can be written as a sum over the eigenstates of the $\delta$-Bose gas with $N$ particles,
\begin{equation}
\label{<ZZZ>[psi]}
\left\langle Z(y_{1},t)\ldots Z(y_{N},t)\right\rangle=\sum_{r}\rme^{-tE_{r}}\langle y_{1},\ldots,y_{N}|\psi_{r}\rangle\langle\psi_{r}|0\rangle\;,
\end{equation}
which inserting in (\ref{G1[<ZZZ>]}) yields
\begin{eqnarray}
\label{G1[psi]}
&& G_{1}=1+\sum_{N=1}^{\infty}\frac{(-1)^{N}\rme^{tN/24}}{N!}\sum_{r}\rme^{-tE_{r}}|\psi_{r}(0,\ldots,0)|^{2}\nonumber\\
&&\qquad\qquad\qquad \times\sum_{\ell_{1},\ldots,\ell_{N}=1}^{n}\rme^{-(s_{\ell_{1}}+\ldots+s_{\ell_{N}})}\frac{\psi_{r}(x_{\ell_{1}},\ldots,x_{\ell_{N}})}{\psi_{r}(0,\ldots,0)}\;.
\end{eqnarray}
The required matrix elements are computed in \cite{D10.2,CC07.1,CC07.2}. We follow mostly the notation in \cite{PS10.1}. Let us pick a number of clusters of particles $M$, $1\leq M\leq N$, and the positive integers $n_{\alpha}$, $\alpha=1,\ldots,M$ counting the number of particles in each cluster, such that
\begin{equation}
\sum_{\alpha=1}^{M}n_{\alpha}=N\;.
\end{equation}
We also introduce $M$ real momenta $q_{\alpha}$, $\alpha=1,\ldots,M$, and set $\boldsymbol{q}=(q_{1},\ldots,q_{M})$, $\boldsymbol{n}=(n_{1},\ldots,n_{M})$, $\boldsymbol{y}=(y_{1},\ldots,y_{N})$. Then the eigenvectors of the $N$-particle Hamiltonian $H_{N}$ are indexed by $M$, $\boldsymbol{q}$, $\boldsymbol{n}$ and
\begin{equation}
\fl\qquad \psi_{\boldsymbol{q},\boldsymbol{n}}^{(M)}(\boldsymbol{y}) = {\sum_{p\in\mathcal{P}}}'A_{p}(\boldsymbol{y})\exp\Bigg[\rmi\sum_{\alpha=1}^{M}q_{\alpha}\sum_{c\in\Omega_{\alpha}(p)}y_{c}-\frac{1}{4}\sum_{\alpha=1}^{M}\sum_{c,c'\in\Omega_{\alpha}(p)}|y_{c}-y_{c'}|\Bigg]\;,
\end{equation}
see \cite{PS10.1}, Eq. (3.17). $\mathcal{P}$ is the permutation group with $N$ elements. The prime means that the summation is only over the subset of $\mathcal{P}$ consisting of permutations between distinct clusters. The coefficients $A_{p}(\boldsymbol{y})$ and the sets $\Omega_{\alpha}(p)$ are defined in \cite{PS10.1}, Section III.B, where for the present purpose one only needs to know that
\begin{equation}
\label{sum p Ap}
{\sum_{p\in\mathcal{P}}}'A_{p}(\boldsymbol{y})=\psi_{\boldsymbol{q},\boldsymbol{n}}^{(M)}(0)\;
\end{equation}
and that the sets $\Omega_{\alpha}(p)$, $\alpha=1,\ldots,M$ form a partition of the set of integers $1,\ldots,N$. We now write explicitly the sum over the $\ell_{a}$, $a=1,\ldots,N$, in (\ref{G1[psi]}),
\begin{equation}
\fl\qquad \sum_{\ell_{1},\ldots,\ell_{N}=1}^{n}\rme^{-(s_{\ell_{1}}+\ldots+s_{\ell_{N}})}\psi_{\boldsymbol{q},\boldsymbol{n}}^{(M)}(x_{\ell_{1}},\ldots,x_{\ell_{N}})={\sum_{p\in\mathcal{P}}}'A_{p}(x_{\ell_{1}},\ldots,x_{\ell_{N}})\rme^{\phi(\boldsymbol{\ell},p)}\;
\end{equation}
with $\boldsymbol{\ell}=(\ell_{1},\ldots,\ell_{N})$. The phase $\phi(\boldsymbol{\ell},p)$ equals
\begin{eqnarray}
&& \phi(\boldsymbol{\ell},p)=\sum_{\alpha=1}^{M}\Bigg[-\sum_{j=1}^{n}s_{j}m_{\alpha,j}(\boldsymbol{\ell},p)+\rmi q_{\alpha}\sum_{j=1}^{n}x_{j}m_{\alpha,j}(\boldsymbol{\ell},p)\nonumber\\
&&\qquad\qquad\qquad\qquad -\frac{1}{4}\sum_{j,k=1}^{n}|x_{j}-x_{k}|m_{\alpha,j}(\boldsymbol{\ell},p)m_{\alpha,k}(\boldsymbol{\ell},p)\Bigg]\;,
\end{eqnarray}
with
\begin{equation}
m_{\alpha,j}(\boldsymbol{\ell},p)=\#\{c\in\Omega_{\alpha}(p),\ell_{c}=j\}\;,\quad j=1,\ldots,n\;.
\end{equation}

As in the simpler case of the two-point function \cite{PS10.1}, the sum over $p$ and the sum over $\boldsymbol{\ell}$ are coupled and it does not seem possible to simplify our expression. In order to push forward the calculation, we now assume that these two sums factorize. While this factorization is not exact, the fact that we recover the $n$-point distribution of the Airy process strongly suggests that the factorization assumption becomes a valid approximation in the long time limit. Thus for fixed $p$ we sum the exponential of the phase over $\boldsymbol{\ell}$. This sum does not depend on $p$ and (\ref{sum p Ap}) can be employed to arrive at 
\begin{eqnarray}
\label{3.11}
\fl&&\quad \sum_{\boldsymbol{\ell}}\rme^{-(s_{\ell_{1}}+\ldots+s_{\ell_{N}})}\psi_{\boldsymbol{q},\boldsymbol{n}}^{(M)}(x_{\ell_{1}},\ldots,x_{\ell_{N}})/\psi_{\boldsymbol{q},\boldsymbol{n}}^{(M)}(0)\nonumber\\
\fl&&\quad \simeq\sum_{\boldsymbol{\ell}}\prod_{\alpha=1}^{M}\exp\Bigg[\sum_{j=1}^{n}(-s_{j}+\rmi q_{\alpha}x_{j})m_{\alpha,j}(\boldsymbol{\ell})-\frac{1}{4}\sum_{j,k=1}^{n}|x_{j}-x_{k}|m_{\alpha,j}(\boldsymbol{\ell})m_{\alpha,k}(\boldsymbol{\ell})\Bigg]\;.
\end{eqnarray}
We use the identity
\begin{equation}
\rme^{au+bv+cuv}=\rme^{c\partial_{a}\partial_{b}}\rme^{au+bv}\;,
\end{equation}
which can be proved by expanding both sides as formal series in $c$, and obtain
\begin{eqnarray}
\fl&&\hspace{40pt}\sum_{\boldsymbol{\ell}}\prod_{\alpha=1}^{M}\exp\Bigg[\sum_{j=1}^{n}(-s_{j}+\rmi q_{\alpha}x_{j})m_{\alpha,j}(\boldsymbol{\ell})-\frac{1}{4}\sum_{j,k=1}^{n}|x_{j}-x_{k}|m_{\alpha,j}(\boldsymbol{\ell})m_{\alpha,k}(\boldsymbol{\ell})\Bigg] \nonumber\\
\fl&&\hspace{90pt}= \sum_{\boldsymbol{\ell}}\prod_{\alpha=1}^{M}\rme^{-\frac{1}{4}\sum\limits_{j,k=1}^{n}|x_{j}-x_{k}|\partial_{s_{j}}\partial_{s_{k}}}\rme^{\sum\limits_{j=1}^{n}(-s_{j}+\rmi q_{\alpha}x_{j})m_{\alpha,j}(\boldsymbol{\ell})}\nonumber\\
\fl&&\hspace{90pt} =\prod_{\alpha=1}^{M}\rme^{-\frac{1}{4}\sum\limits_{j,k=1}^{n}|x_{j}-x_{k}|\partial_{s_{j}}\partial_{s_{k}}}\Bigg(\sum_{\ell=1}^{n}\rme^{-s_{\ell}+\rmi q_{\alpha}x_{\ell}}\Bigg)^{n_{\alpha}}\;.
\end{eqnarray}

%%%%%%%%%%%%%%%%%%%%%%%%%%%%%%%%%%%%%
%%%%%%%%%%%%%%%%     %%%%%%%%%%%%%%%%
%%%%%%%%%%%%%%%%  4  %%%%%%%%%%%%%%%%
%%%%%%%%%%%%%%%%     %%%%%%%%%%%%%%%%
%%%%%%%%%%%%%%%%%%%%%%%%%%%%%%%%%%%%%

\section{\texorpdfstring{$n$}{n}-point generating function}
\label{Section Fredholm G1}
\setcounter{equation}{0}
\subsection{Fredholm determinant}
The generating function $G_{1}$ with factorization (\ref{3.11}) will be denoted by $G_{1}^{\sharp}$. The required normalization of the eigenfunctions are computed in \cite{CC07.1,CC07.2,D10.2}. According to \cite{PS10.1}, Eq. (4.5), it holds
\begin{equation}
|\psi_{\boldsymbol{q},\boldsymbol{n}}^{(M)}(0,\ldots,0)|^{2}=N!\det\Bigg(\frac{1}{\frac{1}{2}(n_{\alpha}+n_{\beta})+\rmi(q_{\alpha}-q_{\beta})}\Bigg)_{\alpha,\beta=1,\ldots,M}\;.
\end{equation}
In other contexts, determinantal formulas for such normalizations have also been found \cite{G83.1,KK88.1}. The energy of the eigenstate indexed by $M$, $\boldsymbol{q}$, $\boldsymbol{n}$ is given by
\begin{equation}
E_{\boldsymbol{q},\boldsymbol{n}}^{(M)}=\frac{1}{2}\sum_{\alpha=1}^{M}n_{\alpha}q_{\alpha}^{2}-\frac{1}{24}\sum_{\alpha=1}^{M}(n_{\alpha}^{3}-n_{\alpha})\;,
\end{equation}
and the properly normalized sum over the eigenstates reads
\begin{equation}
\sum_{r}\equiv\sum_{M=1}^{\infty}\frac{1}{M!}\prod_{\alpha=1}^{M}\Bigg(\int_{-\infty}^{\infty}\frac{\rmd q_{\alpha}}{2\pi}\sum_{n_{\alpha}=1}^{\infty}\Bigg)\openone_{\big\{N=\sum\limits_{\alpha=1}^M n_{\alpha}\big\}}\;.
\end{equation}
Then $G_{1}^{\sharp}$ rewrites as
\begin{eqnarray}
\label{G1[M,qj,nj]}
&&\hspace{-64pt} G_{1}^{\sharp}=1+\sum_{M=1}^{\infty}\frac{1}{M!}\prod_{\alpha=1}^{M}\Bigg(\int_{-\infty}^{\infty}\frac{\rmd q_{\alpha}}{2\pi}\sum_{n_{\alpha}=1}^{\infty}\Bigg)\det\Bigg(\frac{1}{\frac{1}{2}(n_{\alpha}+n_{\beta})+\rmi(q_{\alpha}-q_{\beta})}\Bigg)_{\alpha,\beta=1,\ldots,M}\nonumber\\
&&\hspace{-60pt} \times\prod_{\alpha=1}^{M}\exp\Big((tn_{\alpha}^{3}/24) -\frac{1}{4}\sum\limits_{j,k=1}^{n}|x_{j}-x_{k}|\partial_{s_{j}}\partial_{s_{k}}\Big)\Bigg[-\rme^{-tq_{\alpha}^{2}/2}\Bigg(\sum_{j=1}^{n}\rme^{\rmi x_{j}q_{\alpha}-s_{j}}\Bigg)\Bigg]^{n_{\alpha}}.
\end{eqnarray}
One observes that $G_{1}^{\sharp}$ is equal to a Fredholm determinant,
\begin{equation}
G_{1}^\sharp=\det(\openone+R)\;,
\end{equation}
where $R$ has the kernel
\begin{eqnarray}\label{kernel R}
&&\hspace{-64pt} R(q,m;q',m')\nonumber\\
&&\hspace{-44pt} =\frac{1}{2\pi}\,\frac{\exp\Big((tm^{3}/24) -\frac{1}{4}\sum\limits_{j,k=1}^{n}|x_{j}-x_{k}|\partial_{s_{j}}\partial_{s_{k}}\Big)\Bigg[-\rme^{-\frac{t}{2}q^{2}}\Bigg(\sum\limits_{j=1}^{n}\rme^{\rmi x_{j}q-s_{j}}\Bigg)\Bigg]^{m}}{\frac{1}{2}(m+m')+\rmi(q-q')}\;,
\end{eqnarray}
compare with Eq. (4.9) of \cite{PS10.1} in the two-point case.

We want to perform the integration over the $q_{\alpha}$ and the summation over the $n_{\alpha}$ inside the Fredholm determinant. For this we first use
\begin{equation}
\frac{1}{\frac{1}{2}(m+m')+\rmi(q-q')}=\int_{0}^{\infty}\rmd z\,\rme^{-z\left[\frac{1}{2}(m+m')+\rmi(q-q')\right]}\;.
\end{equation}
Then the kernel $R$ can be written as a product of two operators,
\begin{equation}
R(q,m;q',m')=\int_{-\infty}^{\infty}\rmd z\,R_{1}(q,m;z)R_{2}(z;q',m')\;,
\end{equation}
with
\begin{eqnarray}
&&\fl\qquad\qquad R_{1}(q,m;z)=\openone_{\{z>0\}}\,\exp\Big(-\rmi qz+(tm^{3}/24)-\frac{1}{4}\sum\limits_{j,k=1}^{n}|x_{j}-x_{k}|\partial_{s_{j}}\partial_{s_{k}}\Big)\nonumber\\
&&\fl\qquad\qquad\qquad\qquad\qquad\qquad\qquad\qquad \times\Bigg[-\rme^{-\frac{1}{2}z}\rme^{-tq^{2}/2}\Bigg(\sum_{j=1}^{n}\rme^{\rmi x_{j}q-s_{j}}\Bigg)\Bigg]^{m}\;
\end{eqnarray}
and
\begin{equation}
R_{2}(z;q',m')=\openone_{\{z>0\}}\,\frac{1}{2\pi}\,\rme^{-\frac{1}{2}m'z}\rme^{\rmi q'z}\;.
\end{equation}
Since $\det(\openone+R_{1}R_{2})=\det(\openone+R_{2}R_{1})$, one arrives at
\begin{equation}
G_{1}^{\sharp}=\det(\openone+N)\;,
\end{equation}
where
\begin{equation}
\label{kernel N}
N(z,z')=\int_{-\infty}^{\infty}\rmd q\,\sum_{m=1}^{\infty}R_{2}(z;q,m)R_{1}(q,m;z')\;.
\end{equation}
More explicitly, it holds
\begin{eqnarray}
\label{kernel N}
&&\fl\qquad\qquad N(z,z')=\openone_{\{z,z'>0\}}\,\exp\Big(-\frac{1}{4}\sum\limits_{j,k=1}^{n}|x_{j}-x_{k}|\partial_{s_{j}}\partial_{s_{k}}\Big)\nonumber\\
&&\fl\qquad\qquad \times\int_{-\infty}^{\infty}\frac{\rmd q}{2\pi}\sum_{m=1}^{\infty}\rme^{\rmi q(z-z')}\rme^{tm^{3}/24}\Bigg[-\rme^{-\frac{1}{2}(z+z')}\rme^{-tq^{2}/2}\Bigg(\sum_{j=1}^{n}\rme^{\rmi x_{j}q-s_{j}}\Bigg)\Bigg]^{m}\;.
\end{eqnarray}
We now use the identity
\begin{equation}
\rme^{tm^{3}/24}=\int_{-\infty}^{\infty}\rmd u\,\Ai(u)\rme^{(t/8)^{1/3}mu}\;,
\end{equation}
in order to perform the summation over $m$. Defining the function
\begin{equation}
\Phi(u_{1},\ldots,u_{n})=\frac{\rme^{u_{1}}+\ldots+\rme^{u_{n}}}{1+\rme^{u_{1}}+\ldots+\rme^{u_{n}}}\;,
\end{equation}
one has
\begin{eqnarray}
&&\fl\qquad N(z,z')=-\openone_{\{z,z'>0\}}\,\exp\Big(-\frac{1}{4}\sum\limits_{j,k=1}^{n}|x_{j}-x_{k}|\partial_{s_{j}}\partial_{s_{k}}\Big)\int_{-\infty}^{\infty}\frac{\rmd q\,\rmd u}{2\pi}\Ai(u)\rme^{\rmi q(z-z')}\nonumber\\
&&\fl\qquad\qquad\qquad \times\Phi\left((t/8)^{1/3}u-\tfrac{1}{2}(z+z')-tq^{2}/2+\rmi x_{j}q-s_{j},\,j=1,\ldots,n\right)\;.
\end{eqnarray}

Let us introduce the parameter $\alpha=(t/2)^{1/3}$. We remove the terms $\rmi x_{j}q$ from $\Phi$ by using
\begin{equation}
\label{exp shift}
f(z+a)=\exp[a\partial_{z}]f(z)
\end{equation}
and shift $u$ by $u\to u+2^{2/3}\alpha^{2}q^{2}+2^{-1/3}\alpha^{-1}(z+z')$ to obtain
\begin{eqnarray}
&&\hspace{-66pt}N(z,z')=-\openone_{\{z,z'>0\}}\,\exp\Big(-\frac{1}{4}\sum\limits_{j,k=1}^{n}|x_{j}-x_{k}|\partial_{s_{j}}\partial_{s_{k}}\Big)\int_{-\infty}^{\infty}\frac{\rmd q\,\rmd u}{2\pi}\Ai(u+2^{2/3}\alpha^{2}q^{2}\nonumber\\
&&\hspace{-66pt} +2^{-1/3}\alpha^{-1}(z+z'))\exp\Big(\rmi q(z-z'-\sum\limits_{j=1}^{n}x_{j}\partial_{s_{j}})\Big)\Phi\left(2^{-2/3}\alpha u-s_{j},\,j=1,\ldots,n\right).
\end{eqnarray}
Integrating over $q$,
\begin{eqnarray}
&&\fl\qquad\qquad \int_{-\infty}^{\infty}\frac{\rmd q}{2\pi}\Ai(aq^{2}+b)\rme^{\rmi cq}\nonumber\\
&&\fl\qquad\qquad =2^{-1/3}a^{-1/2}\Ai\left(2^{-2/3}\left(b+a^{-1/2}c\right)\right)\Ai\left(2^{-2/3}\left(b-a^{-1/2}c\right)\right)\;,
\end{eqnarray}
one arrives at
\begin{eqnarray}
&&\hspace{-60pt} N(z,z')=-\openone_{\{z,z'>0\}}\,\exp\Big(-\frac{1}{4}\sum\limits_{j,k=1}^{n}|x_{j}-x_{k}|\partial_{s_{j}}\partial_{s_{k}}\Big)\int_{-\infty}^{\infty}\rmd u\,2^{-2/3}\alpha^{-1}\nonumber\\
&&\hspace{-46pt} \times\Ai\Big(2^{-2/3}u+\alpha^{-1}z-2^{-1}\alpha^{-1}\sum_{j=1}^{n}x_{j}\partial_{s_{j}}\Big)\Ai\Big(2^{-2/3}u+\alpha^{-1}z'+2^{-1}\alpha^{-1}\sum_{j=1}^{n}x_{j}\partial_{s_{j}}\Big)\nonumber\\
&&\hspace{-46pt} \times\Phi\left(2^{-2/3}\alpha u-s_{j},\,j=1,\ldots,n\right)\;.
\end{eqnarray}
Applying (\ref{exp shift}) to extract the operator $2^{-1}\alpha^{-1}\sum_{j=1}^{n}x_{j}\partial_{s_{j}}$ from the Airy functions and performing the changes of variables $u\to2^{2/3}u$ and $z\to\alpha z$, we obtain
\begin{eqnarray}
&&\hspace{-66pt} \tilde{N}(z,z')=-\alpha N(\alpha z,\alpha z')=\openone_{\{z,z'>0\}}\,\exp\Big(-\frac{1}{4}\sum\limits_{j,k=1}^{n}|x_{j}-x_{k}|\partial_{s_{j}}\partial_{s_{k}}\\
&&\hspace{-62pt}-(2\alpha)^{-1}(\partial_{z}-\partial_{z'})\sum\limits_{j=1}^{n}x_{j}\partial_{s_{j}}\Big)\int_{-\infty}^{\infty}\rmd u\,\Ai(u+z)\Ai(u+z')\Phi(\alpha u-s_{1},\ldots,\alpha u-s_{n})\;,\nonumber
\end{eqnarray}
compare with Eq. (4.19) of \cite{PS10.1}. With these rearrangements the generating function $G_{1}^\sharp$ is written as
\begin{equation}
G_{1}^{\sharp}=\det(\openone-\tilde{N})\;.
\end{equation}

We return to the original generating function (under the factorization assumption)
\begin{equation}
G^{\sharp}(\boldsymbol{s};\boldsymbol{x},t) = G_{1}^{\sharp}\left(s_{1}-\tfrac{1}{2t}x_{1}^{2},\ldots,s_{n}-\tfrac{1}{2t}x_{n}^{2};\boldsymbol{x},t\right)\;.
\end{equation}
Applying (\ref{exp shift}) to subtract the parabolic profile, we arrive at
\begin{equation}
\label{Gsharp[L]}
G^{\sharp}=\det(\openone-L)
\end{equation}
with
\begin{eqnarray}
\label{kernel L}
&&\fl\qquad L(z,z')=\openone_{\{z,z'>0\}}\,\exp\Big(-\frac{1}{4}\sum\limits_{j,k=1}^{n}|x_{j}-x_{k}|\partial_{s_{j}}\partial_{s_{k}}-\frac{1}{4\alpha^{3}}\sum\limits_{j=1}^{n}x_{j}^{2}\partial_{s_{j}}-(2\alpha)^{-1}(\partial_{z}-\partial_{z'})\nonumber\\
&&\fl\qquad\qquad\qquad \times\sum\limits_{j=1}^{n}x_{j}\partial_{s_{j}}\Big)\int_{-\infty}^{\infty}\rmd u\,\Ai(u+z)\Ai(u+z')\Phi(\alpha u-s_{1},\ldots,\alpha u-s_{n})\;.
\end{eqnarray}

\subsection{Rewriting of the kernel \texorpdfstring{$L$}{L} in terms of the Airy Hamiltonian}
We will now rewrite the kernel $L$ in terms of the Airy Hamiltonian $H$. In order to eliminate the derivatives with respect to the $s_{j}$, we first write the function $\Phi$ in Fourier form as
\begin{equation}
\fl\qquad \Phi(\alpha u-s_{1},\ldots,\alpha u-s_{n})=\int_{-\infty}^{\infty}\rmd r_{1}\,\ldots\rmd r_{n}\,\rme^{-\rmi\sum\limits_{j=1}^{n}r_{j}(\alpha u-s_{j})}\hat{\Phi}(r_{1},\ldots,r_{n})\;.
\end{equation}
Then, setting $\boldsymbol{r}=(r_{1},\ldots,r_{n})$, the kernel $L$ takes the form
\begin{equation}
\label{L[Lhat]}
L(z,z')=\int_{-\infty}^{\infty}\rmd r_{1}\,\ldots\rmd r_{n}\,\exp\Bigg(\rmi\sum_{j=1}^{n}r_{j}s_{j}\Bigg)\hat{L}(\boldsymbol{r})\hat{\Phi}(\boldsymbol{r})\;,
\end{equation}
where
\begin{eqnarray}
&&\fl\qquad \hat{L}(\boldsymbol{r})=\openone_{\{z,z'>0\}}\,\exp\Bigg(\frac{1}{4}\sum_{j,k=1}^{n}|x_{j}-x_{k}|r_{j}r_{k}-\frac{\rmi}{4\alpha^{3}}\sum_{j=1}^{n}x_{j}^{2}r_{j}-\frac{\rmi}{2\alpha}(\partial_{z}-\partial_{z'})\nonumber\\
&&\fl\qquad\qquad\qquad\quad \times\sum_{j=1}^{n}x_{j}r_{j}\Bigg)\int_{-\infty}^{\infty}\rmd u\,\exp\Bigg(-\rmi\alpha u\sum_{j=1}^{n}r_{j}\Bigg)\Ai(u+z)\Ai(u+z')\;.
\end{eqnarray}
Since the rest of the argument is independent of the particular function $\Phi$, we will work only on $\hat{L}$. We use (\ref{exp shift}) for the derivatives with respect to $z$ and $z'$ and shift $u$ by $\rmi(2\alpha)^{-1}\sum_{j=1}^{n}x_{j}r_{j}$ in the integral. Taking into account the fact that $x_{1}<\ldots<x_{n}$, one obtains
\begin{eqnarray}
&&\fl\qquad \hat{L}(\boldsymbol{r})=\openone_{\{z,z'>0\}}\,\exp\Bigg(\sum_{j<k}x_{k}r_{j}r_{k}\Bigg)\exp\Bigg(\frac{1}{2}\sum_{j=1}^{n}x_{j}r_{j}^{2}\Bigg)\exp\Bigg(-\frac{\rmi}{4\alpha^{3}}\sum_{j=1}^{n}x_{j}^{2}r_{j}\Bigg)\nonumber\\
&&\fl\qquad\qquad\quad \times\int_{-\infty}^{\infty}\rmd u\,\exp\Bigg(-\rmi\alpha u\sum\limits_{j=1}^{n}r_{j}\Bigg)\Ai(u+z)\Ai\Bigg(u+z'+\frac{\rmi}{\alpha}\sum_{j=1}^{n}x_{j}r_{j}\Bigg)\;.
\end{eqnarray}
We introduce the operators $H$, $U$ and $D$, defined by
\begin{equation}
H=-\partial_{u}^{2}+u\;,\qquad U=u\;,\qquad D=\partial_{u}\;,
\end{equation}
and the function $\Ai_{z}(u)=\Ai(u+z)$, which is the eigenfunction of the Airy Hamiltonian $H$ with eigenvalue $-z$. The Airy kernel $K$ is the projection on the negative eigenstates of $H$, \textit{i.e.} $K|\Ai_{z}\rangle=\openone_{\{z>0\}}|\Ai_{z}\rangle$. Thus, $\hat{L}(\boldsymbol{r})$ can be written as
\begin{eqnarray}
\label{Lhat 1}
&&\fl\qquad \hat{L}(\boldsymbol{r})=\exp\Bigg(\sum_{j<k}x_{k}r_{j}r_{k}\Bigg)\exp\Bigg(\frac{1}{2}\sum\limits_{j=1}^{n}x_{j}r_{j}^{2}\Bigg)\exp\Bigg(-\frac{\rmi}{4\alpha^{3}}\sum\limits_{j=1}^{n}x_{j}^{2}r_{j}\Bigg)\nonumber\\
&&\fl\qquad\qquad\quad \times\langle\Ai_{z'}|K\,\exp\Bigg(\frac{\rmi}{\alpha}\sum\limits_{j=1}^{n}x_{j}r_{j}D\Bigg)\exp\Bigg(-\rmi\alpha\sum\limits_{j=1}^{n}r_{j}U\Bigg)K|\Ai_{z}\rangle\;.
\end{eqnarray}
This expression for $\hat{L}(\boldsymbol{r})$ will now be rearranged using the following lemma.
\begin{lemma}
\label{Lemma HUD}
Let $\lambda_{j}\geq0$, $j=1,\ldots,n+1$ and $\mu_{k}\in\rmi\mathbb{R}$, $k=1,\ldots,n$. Then the following relation holds:
\begin{eqnarray}
&& \rme^{\lambda_{1}H}\rme^{\mu_{1}U}\rme^{\lambda_{2}H}\rme^{\mu_{2}U}\ldots\rme^{\lambda_{n}H}\rme^{\mu_{n}U}\rme^{\lambda_{n+1}H}=\exp\Bigg(\sum_{j=1}^{n}(\lambda_{1}+\ldots+\lambda_{j})^{2}\mu_{j}\Bigg)\nonumber\\
&&\quad \times\exp\Bigg(\sum_{j=1}^{n}(\lambda_{1}+\ldots+\lambda_{j})\mu_{j}^{2}\Bigg)\exp\Bigg(2\!\!\!\sum_{1\leq j<k\leq n}\!\!\!\mu_{j}\mu_{k}(\lambda_{1}+\ldots+\lambda_{k})\Bigg)\nonumber\\
&&\quad \times\exp\Bigg(-2\sum_{j=1}^{n}(\lambda_{1}+\ldots+\lambda_{j})\mu_{j}D\Bigg)\exp\Bigg(\sum_{j=1}^{n}\mu_{j}U\Bigg)\nonumber\\
&&\quad \times\exp\Big((\lambda_{1}+\ldots+\lambda_{n+1})H\Big)\;.
\end{eqnarray}
\end{lemma}
\begin{proof}
The operators $H$, $U$ and $D$ satisfy the commutation relations
\begin{equation}
[H,U]=-2D\;,\qquad [H,D]=-1\;,\qquad [U,D]=-1\;.
\end{equation}
Then, the Baker-Campbell-Hausdorff formula yields the commutation relations for the exponentials. One has
\begin{eqnarray}
&& \rme^{\lambda H}\rme^{\mu U}=\rme^{\lambda^{2}\mu}\rme^{\lambda\mu^{2}}\rme^{-2\lambda\mu D}\rme^{\mu U}\rme^{\lambda H}\\
&& \rme^{\alpha U}\rme^{\beta D}=\rme^{-\alpha\beta}\rme^{\beta D}\rme^{\alpha U}\\
&& \rme^{\alpha H}\rme^{\beta D}=\rme^{-\alpha\beta}\rme^{\beta D}\rme^{\alpha H}\;.
\end{eqnarray}
Using iteratively these relations, the proof of the lemma is straightforward.
\end{proof}
Lemma \ref{Lemma HUD} with $\lambda_{j}=(x_{j-1}-x_{j})/2\alpha^{2}$ and $\mu_{j}=-\rmi\alpha r_{j}$ ($x_{0}=x_{n+1}=0$) allows to rewrite the expression (\ref{Lhat 1}) of $\hat{L}(\boldsymbol{r})$ as
\begin{eqnarray}
&& \hat{L}(\boldsymbol{r})=\langle\Ai_{z'}|K\rme^{(-x_{1}/2\alpha^{2})H}\rme^{-\rmi\alpha r_{1}U}\rme^{((x_{1}-x_{2})/2\alpha^{2})H}\rme^{-\rmi\alpha r_{2}U}\nonumber\\
&&\qquad\qquad\quad \times\ldots\rme^{((x_{n-1}-x_{n})/2\alpha^{2})H}\rme^{-\rmi\alpha r_{n}U}\rme^{(x_{n}/2\alpha^{2})H}K|\Ai_{z}\rangle\;.
\end{eqnarray}
Inserting in (\ref{L[Lhat]}), one finds the following expression for the kernel $L$
\begin{eqnarray}
&&\fl\quad L(z,z')=\int_{-\infty}^{\infty}\rmd u_{1}\,\ldots\,\rmd u_{n}\,\langle\Ai_{z'}|K\rme^{(-x_{1}/2\alpha^{2})H}|u_{1}\rangle\langle u_{1}|\rme^{((x_{1}-x_{2})/2\alpha^{2})H}|u_{2}\rangle\times\ldots\nonumber\\
&&\fl\quad \times\langle u_{n-1}|\rme^{((x_{n-1}-x_{n})/2\alpha^{2})H}|u_{n}\rangle\langle u_{n}|\rme^{(x_{n}/2\alpha^{2})H}K|\Ai_{z}\rangle\Phi(\alpha u_{1}-s_{1},\ldots,\alpha u_{n}-s_{n})\;.
\end{eqnarray}
A similarity transformation plus the fact that $K$ is a projection, $K^{2}=K$, finally gives the result announced in Eq. (\ref{Gsharp[Q]}).

%%%%%%%%%%%%%%%%%%%%%%%%%%%%%%%%%%%%%
%%%%%%%%%%%%%%%%     %%%%%%%%%%%%%%%%
%%%%%%%%%%%%%%%%  5  %%%%%%%%%%%%%%%%
%%%%%%%%%%%%%%%%     %%%%%%%%%%%%%%%%
%%%%%%%%%%%%%%%%%%%%%%%%%%%%%%%%%%%%%

\section{Conclusions}
\setcounter{equation}{0}
Following the strategy in \cite{PS10.1}, we extended our results to the spatial $n$-point generating function at some common time $t$. We also simplified the algebra used for the $2$-point function.

The decoupling assumption remains somewhat mysterious. For small $M$ we checked that (\ref{3.11}) is not an identity. Of course, there could be cancellations such that (\ref{Gsharp[L]}) survives as an identity for $G$. We have no idea how to check such a claim directly. On the other hand, the long time limit of $G^{\sharp}$ agrees with the conjecture based on universality, which is counterintuitive since usually approximations become amplified as $t\to\infty$.

Our results hold only for sharp wedge initial data. There are two further initial conditions for which one would like to solve the KPZ equation. One is the flat initial condition, $h(x,0)=0$. Then
\begin{equation}
h(0,t)=\log\int\rmd x\,Z(x,t)\;.
\end{equation}
The second one is stationary initial condition. This corresponds to solving (\ref{Z}) with two-sided Brownian motion as initial condition, \textit{i.e.}
\begin{equation}
Z(\pm x,0)=\rme^{b_{\pm}(x)}\;,\quad x\geq0\;,
\end{equation}
with $b_{+}$ and $b_{-}$ two independent standard Brownian motions. Even at the level of the one-point function, both cases remain as a challenge.

%%%%%%%%%%%%%%%%%%%%%%%%%%%%%%%%%%%%%
%%%%%%%%%%%%%%%%     %%%%%%%%%%%%%%%%
%%%%%%%%%%%%%%%%  A  %%%%%%%%%%%%%%%%
%%%%%%%%%%%%%%%%     %%%%%%%%%%%%%%%%
%%%%%%%%%%%%%%%%%%%%%%%%%%%%%%%%%%%%%

\begin{appendix}
\section{Matrix kernel and scalar kernel}
\label{Appendix matrix kernel}
\setcounter{equation}{0}
For better readability we use sans-serif for operators on $L^2(\mathbb{R})\otimes\mathbb{C}^{n}$, while the matrix elements, as $\mathsf{K}_{ij}$, $i,j = 1,\ldots,n$, are operators on $L^2(\mathbb{R})$. The operator $\mathsf{K}$ of Eq. (\ref{matrix kernel}) can be written as
\begin{equation}
\mathsf{K}=\mathsf{P}(\mathsf{T}^{-}\mathsf{K}^0+\mathsf{T}^{+}(\mathsf{K}^0-1))\mathsf{P}\;,
\end{equation}
where 
\begin{equation}
\mathsf{K}^{0}_{ij}=\delta_{ij}K\;,\quad \mathsf{P}_{ij}=\delta_{ij}P_{a_{j}}\;,
\end{equation}
and $\mathsf{T}^{-}$, $\mathsf{T}^{+}$ are lower triangular, resp. strictly upper triangular, according to
\begin{equation}
\mathsf{T}^{-}_{ij} = \openone_{\{i\geq j\}}\rme^{(w_{i}-w_{j})H} \;, \quad
\mathsf{T}^{+}_{ij}=\openone_{\{i < j\}}\rme^{(w_{i}-w_{j})H}\;.
\end{equation}
Hence the Fredholm determinant of the operator $\mathsf{K}$ can be rewritten as
\begin{equation}
\det(1-\mathsf{K})=\det(1+\mathsf{P}\mathsf{T}^{+})\det(1-(1+\mathsf{P}\mathsf{T}^{+})^{-1}\mathsf{P}(\mathsf{T}^{-}+\mathsf{T}^{+})\mathsf{K})\;.
\end{equation}
Recall that
\begin{equation}
[H,K]=0\;, \quad K^{2}=K\;, \quad P_{a}^{2}=P_{a}\;.
\end{equation}
For any integer $m\geq1$, $\tr(\mathsf{P}\mathsf{T}^{+})^{m}=0$, which implies that $\det(1+\mathsf{P}\mathsf{T}^{+})=1$. Since $\mathsf{P}\mathsf{T}^{+}$ is nilpotent, the expansion of $(1+\mathsf{P}\mathsf{T}^{+})^{-1}$ terminates as
\begin{equation}
(1+\mathsf{P}\mathsf{T}^{+})^{-1}=1-\mathsf{P}\mathsf{T}^{+}+\ldots+(-1)^{n-1}(\mathsf{P}\mathsf{T}^{+})^{n-1}\;.
\end{equation}
We now consider the operator
\begin{equation}
\mathsf{B}=(1+\mathsf{T}^{+})(1+\mathsf{P}\mathsf{T}^{+})^{-1}\mathsf{P}(\mathsf{T}^{-}+\mathsf{T}^{+})(1+\mathsf{T}^{+})^{-1}\mathsf{K}^0\;.
\end{equation}
Since $1+\mathsf{T}^{+}$ and $\mathsf{K}^0$ commute, one has $\det(1-\mathsf{K})=\det(1-\mathsf{B})$. The inverse of $1+\mathsf{T}^{+}$ is given by
\begin{equation}
((1+\mathsf{T}^{+})^{-1})_{ij}= \delta_{ij}1 - \delta_{ij-1}\rme^{(w_{i}-w_{i+1})H}\;.
\end{equation}
Multiplying on the left by $\mathsf{T}^{-}+\mathsf{T}^{+}$ leads to
\begin{equation}
((\mathsf{T}^{-}+\mathsf{T}^{+})(1+\mathsf{T}^{+})^{-1})_{ij}=\delta_{1j}\ \rme^{(w_{i}-w_{1})H} \;.
\end{equation}
Finally, setting $\bar{P}_{a}=1-P_{a}$, one arrives at
\begin{eqnarray}
&&\hspace{0pt}\mathsf{B}_{ij}=\delta_{1j}\big(\rme^{(w_{i}-w_{1})H}K-\bar{P}_{a_{i}}\rme^{(w_{i}-w_{i+1})H}\bar{P}_{a_{i+1}}\rme^{(w_{i+1}-w_{i+2})H}\nonumber\\
&&\hspace{50pt}\times\cdots\times \rme^{(w_{n-1}-w_{n})H}\bar{P}_{a_{n}}\rme^{(w_{n}-w_{1})H}K\big) \;.
\end{eqnarray}
Thus $\det(1-\mathsf{K})=\det(1-\mathsf{B}_{11})$, which is the claimed identity.
\end{appendix}

%%%%%%%%%%%%%%%%%%%%%%%%%%%%%%%%%%%%%
%%%%%%%%%%%%%%%%     %%%%%%%%%%%%%%%%
%%%%%%%%%%%%%%%%  R  %%%%%%%%%%%%%%%%
%%%%%%%%%%%%%%%%     %%%%%%%%%%%%%%%%
%%%%%%%%%%%%%%%%%%%%%%%%%%%%%%%%%%%%%

\end{document}